\documentclass[aps,pra,twocolumn,floatfix,longbibliography,showpacs,superscriptaddress]{revtex4-2}

\usepackage{amsmath,amsfonts,amssymb,bm}
\usepackage{dcolumn}
\usepackage{braket}
\usepackage[final]{graphicx}
\usepackage{tabularx}
\usepackage{graphicx}
\usepackage[update,prepend]{epstopdf}

\usepackage{dsfont}
\usepackage{pifont}
\def\dblone{\mathds{1}}



\usepackage{bbold}
\usepackage{physics}

\usepackage[capitalise]{cleveref}

\usepackage{pgfplots}
\usetikzlibrary{pgfplots.groupplots}
\usetikzlibrary{patterns}
\usepackage{tikzscale}
\usepackage{filecontents}

\begin{document}

\title{Anisotropy factor spectra for weakly allowed electronic transitions in chiral ketones}

\author{Leon A. Kerber}
\affiliation{Dahlem Center for Complex Quantum Systems and Fachbereich Physik, Freie Universität Berlin, D-14195 Berlin, Germany}
\author{Oliver Kreuz}
\affiliation{Fachbereich Chemie, Philipps-Universität Marburg, D-35032 Marburg, Germany}
\author{Tom Ring}
\affiliation{Institut für Physik, Universität Kassel, D-34132 Kassel, Germany}
\author{Hendrike Braun}
\affiliation{Institut für Physik, Universität Kassel, D-34132 Kassel, Germany}
\author{Robert Berger}
\affiliation{Fachbereich Chemie, Philipps-Universität Marburg, D-35032 Marburg, Germany}
\author{Daniel M. Reich}
\email{danreich@zedat.fu-berlin.de}
\affiliation{Dahlem Center for Complex Quantum Systems and Fachbereich Physik, Freie Universität Berlin, D-14195 Berlin, Germany}

\begin{abstract}
Quantum chemical calculations of one-photon absorption, electronic circular dichroism and anisotropy factor spectra for the A-band transition of fenchone, camphor and 3-methylcyclopentanone (3MCP) are reported. While the only weakly allowed nature of the transition leads to comparatively large anisotropies, a proper theoretical description of the absorption for such a transition requires to account for non-Condon effects. We present experimental data for the anisotropy of 3MCP in the liquid phase and show that corresponding Herzberg--Teller corrections are critical to reproduce the main experimental features. The results obtained with our comprehensive theoretical model highlight the importance of the vibrational degree of freedom, paving the way for a deeper understanding of the dynamics in electronic circular dichroism.
\end{abstract}

\maketitle

\section{Introduction}

Optically active chiral molecules differ in their absorption of left and right circularly polarized light, this phenomenon is called circular dichroism (CD) \citep{CDnakanishi}. For radiation in the ultraviolet or visible regime the absorption is primarily determined by electronic transitions. In this context the term electronic circular dichroism is used in contrast to, e.g., vibrational circular dichroism in the infrared. While for unpolarized or linearly polarized absorption (ABS) electric dipole transitions constitute the dominant effect, CD is usually very small due to involving both the electric and magnetic transition dipoles.
Circular dichroism of an electronic band is primarily characterised via its so-called rotatory strength $R$, determined by the integral of the difference in absorption over the entire absorption band, and its so-called dipole strength $D$, the corresponding integral of the unpolarized absorption \cite{berova_application_2007}.
The ratio between CD and ABS yields the so-called anisotropy factor \cite{Kuhn1930}, also called $g$-factor or dissymmetry factor \cite{berova_application_2007}.
The anisotropy factor is often considered to be an advantageous observable since it is independent of the sample density and thickness, which can be difficult to characterise in experiments \cite{covington_similarity_2013}.
Furthermore, the effect of the solvate in liquid-phase experiments seems more readily apparent in anisotropy factor spectra compared to CD or ABS spectra \cite{covington_solvation_2016} and peaks can be assigned with more confidence in high-resolution spectra \cite{santoro_high-resolution_2018}.

A particularly favorable class of molecules for the study of circular dichroism are ketones, for which the anisotropy factor of the first singlet electronic excitation band (the A-band) takes on particularly large values in the order of $10^{-1}$ \cite{CDnakanishi}. Most notably, the three ketones fenchone, camphor and 3-methylcyclopentanone (3MCP) have been the subject of various theoretical and experimental studies \cite{pulm_theoretical_1997,CamphorCyclohexane1966,lin_vibronically_2008,feinleib_vapour-phase_1968,boesl_resonance-enhanced_2013,loge_progress_2009,li_linear_2006,urry1968,dekkers_optical_1976,titze_laser_2014,horsch_circular_2011,
breunig_circular_2009,boesl_von_grafenstein_circular_2006,bornschlegl_investigation_2007,ring_self-referencing_2021}.
The origin of the comparatively large anisotropy factors in these systems lies in the only weakly allowed nature of the transition leading to a comparable magnitude of the electric and magnetic transition moments.
However, the nature of such a transition requires a particularly careful theoretical treatment \cite{moffitt_optical_1959,weigang_vibrational_1965}. Most importantly, it is necessary to go beyond the Franck--Condon \cite{franck_elementary_1926,condon_theory_1926,condon_nuclear_1928} (FC) approximation by including Herzberg--Teller \cite{herzberg_schwingungsstruktur_1933} (HT) contributions \cite{CDnakanishi}.
The combination of Franck--Condon and Herzberg--Teller terms is often abbreviated as FCHT.

One early joint experimental and theoretical study for circular dichroism in ketones was performed in Ref.~\cite{pulm_theoretical_1997} for fenchone and camphor. 
However, in their work they only consider vertical transitions, leading to an underestimation of the dipole strength of the A-band.
The CD spectrum of 3MCP with the inclusion of HT contributions has first been theoretically examined in Ref.~\cite{lin_vibronically_2008} in which HT terms were included for both the electric dipole and the magnetic dipole transition moment, with the latter turning out to be negligible for the A-band. The ABS spectrum was not reported in this work. 

Experimentally, both CD and ABS spectra of 3MCP were investigated for absorption in the gas phase \cite{feinleib_vapour-phase_1968,boesl_resonance-enhanced_2013,loge_progress_2009,li_linear_2006,urry1968} as well as the liquid phase \cite{li_linear_2006,urry1968,dekkers_optical_1976}.
In addition to direct measurements on absorption, circular dichroism also presents itself in ion-yield experiments with several corresponding studies on 3MCP \cite{bornschlegl_investigation_2007,boesl_resonance-enhanced_2013,titze_laser_2014,horsch_circular_2011,loge_progress_2009,breunig_circular_2009,
boesl_von_grafenstein_circular_2006,ring_self-referencing_2021}. In the experiment from Ref.~\cite{loge_progress_2009} the anisotropy factor spectrum for conventional absorption and ion yield showed similarities, although the anisotropy factor in ion yield was appreciably smaller. This has been attributed to saturation effects due to high power densities in ion yield experiments \cite{loge_progress_2009,bornschlegl_investigation_2007}.

However, a comprehensive theoretical treatment of the A-band transition in all three ketones with respect to the HT contributions has hitherto been missing to our knowledge.
To amend this, we provide in this work state-of-the-art numerical calculations and compare our results with the experimental literature. We systematically include all relevant Herzberg--Teller contributions and examine their influence on absorption and anisotropy.
For 3MCP, we furthermore study the contributions of the two conformational forms.
We begin by presenting our calculations for rotatory strengths, dipole strengths, and anisotropy factors of the A-band for all three ketones discussed above. Then, we report results for the vibrational substructure in the electronic spectra. 
Finally, we also provide an experimental measurement for the anisotropy factor spectrum in 3MCP in solution and compare our theory with this data as well as previous experimental work.

This paper is organised as follows. Section 2 introduces the theoretical model for circular dichroism, unpolarized absorption and the anisotropy factor of the A-band transition.
Section 3 presents our numerical results, comparing with the literature as well as our own measurements of the anisotropy in 3MCP in the liquid phase. Finally, section 4 concludes.

\section{Methods}

The A-band corresponds to a transition from the electronic ground state $\text{S}_0$, referred to by index $i$ in the following, to the first electronically excited singlet state $\text{S}_1$, referred to by index $f$.
Using the Born--Oppenheimer approximation \cite{Born-Oppenheimer} we write $\ket{i \vb{v}_i} \equiv \ket{i} \otimes \ket{\vb{v}_i}$ and $\ket{f \vb{v}_f} \equiv \ket{f} \otimes \ket{\vb{v}_f}$ for the vibronic levels with $\vb{v}_i$ and $\vb{v}_f$ being the vibrational levels in the electronic states $i$ and $f$, respectively.
Similarly to Ref. \cite{lin_vibronically_2008}, we neglect the rotational degrees of freedom in our work.

We employ SI units throughout this section unless noted otherwise. The electronic part of the electric dipole moment operator is then given by
\begin{equation}
\hat{\vb*{\mu}}=-\sum_{j=1}^{n_\mathrm{e}} e\hat{\vb{q}}_j,
\end{equation}
where $n_\mathrm{e}$ denotes the total number of electrons, $e$ is the elementary charge and $\hat{\vb{q}}_j$ denotes the position operator of the $j$th electron. The electronic part of the magnetic dipole moment operator is given by
\begin{align}
\hat{\vb{m}}=-\sum_{j=1}^{n_e}\frac{e}{2m_\mathrm{e}}\qty(\hat{\vb{q}}_j\times\hat{\vb{p}}_j),
\end{align}
where $m_\mathrm{e}$ is the electron mass and $\hat{\vb{p}}_j$ denotes the linear momentum operator of the $j$th electron. Within the Born--Oppenheimer approximation, the molecular wavefunction can be written as a product of electronic and nuclear parts with the electronic states depending parametrically on the nuclear coordinates. 
As a consequence, the nuclear part of the dipole operators does not contribute to transition moments, because of the orthogonality of the electronic states.
Moreover, since spin-orbit coupling is expected to be small for the ketones we study in our work, the electron spin contribution to the magnetic transition dipole moments is dropped for the present study \cite{bernadotte_origin-independent_2012}.

We now expand the electronic transition matrix elements of $\hat{\vb*{\mu}}$ with respect to the electronic ground state's normal coordinates $\vb{Q}\equiv\left\{Q_r\right\}$ defined relative to the equilibrium position $\vb{Q}=\vb{0}$. For the electronic electric dipole transition moment to first order this yields the expression
\begin{equation}
\vb*{\mu}_{fi}(\vb{Q})\approx\vb*{\mu}_{fi}\qty(\vb{0})+\sum_r \frac{\partial\vb*{\mu}_{fi}}{\partial Q_r}\eval_{\vb{Q}=\vb{0}}Q_r.
\end{equation}

We use the harmonic approximation for the potential energy surfaces of the two electronic states, i.e., all vibrational levels for an electronic state $a$ are considered as tensor products of harmonic oscillator eigenstates $\ket{v_{r,a}}$, where $r\in\qty{1,2,\ldots,3N-6}$ indicates the different vibrational modes.
This means that $\ket{\vb{v}_a}=\ket{v_{1,a}}\otimes\ket{v_{2,a}}\otimes\ldots\otimes\ket{v_{3N-6,a}}$ with $N$ the number of atoms in the molecule.
Finally, we approximate the molecules to be at temperature zero, i.e., the initial vibrational state is the ground state $\ket{\vb{v}_i}=\ket{\vb{0}_i}=\ket{0_i}\otimes\ket{0_i}\otimes\ldots\otimes\ket{0_i}$.

The vibronic matrix elements are then given by
\begin{equation}
\vb*{\mu}_{f\vb{v}_fi\vb{0}_i}\approx\vb*{\mu}_{fi}\qty(\vb{0})\braket{\vb{v}_f}{\vb{0}_i}+\sum_r \frac{\partial\vb*{\mu}_{fi}}{\partial Q_r}\eval_{\vb{Q}=\vb{0}}\mel{\vb{v}_f}{\hat{Q}_r}{\vb{0}_i}.
\label{eq:mu_vibronic}
\end{equation}
The first term contains the Franck--Condon overlap integral $\braket{\vb{v}_f}{\vb{0}_i}$ whereas the second term involves the Herzberg--Teller overlap integrals \cite{herzberg_schwingungsstruktur_1933}.
In the following we refer to the first summand of \cref{eq:mu_vibronic} as Franck--Condon term and to the second summand as Herzberg--Teller term.

The A-band, i.e.~the $\mathrm{S}_0\rightarrow \mathrm{S}_1$ transition, corresponds approximately to a $n\rightarrow\pi^*$ transition in the carbonyl group of the ketones. In a simplified one-particle picture, this is associated with an electron changing from a nonbinding orbital confined to the O atom to an antibonding $\pi^*$ orbital spread over the C and O atoms of the carbonyl group. In the Franck--Condon approximation and when additionally assuming $C_{2v}$ point group symmetry of a ketone that coincides with the local $C_{2v}$ symmetry of the carbonyl group, this transition is electric dipole-forbidden and magnetic dipole-allowed \cite{barron:2004,atkins1997}. When overall point group symmetry is lowered, for instance to chiral $C_2$ or, as in the present cases, to $C_1$, the selection rule for the electric transition dipole term becomes lifted.
Under these circumstances, $\vb*{\mu}_{fi}\qty(\vb{0})$ and thus the Franck--Condon term is very small such that the Herzberg--Teller term becomes important. Conversely, for the magnetic transition dipole moment, the Franck--Condon term is sufficient since the transition is magnetic-dipole allowed \cite{moffitt_optical_1959,CDnakanishi},
\begin{equation}
\vb{m}_{f\vb{v}_fi\vb{0}_i}\approx\vb{m}_{fi}\qty(\vb{0})\braket{\vb{v}_f}{\vb{0}_i}.
\label{eq:m_vibronic}
\end{equation}

Our main goal is to determine anisotropy factors, defined as the ratio of circular dichroism to unpolarized absorption \cite{CDnakanishi}. These quantities can both be investigated in a frequency-dependent manner, highlighting the vibronic structure of the transition in question, or as "band values" obtained via integration over the entire absorption band. From a theoretical point of view, the band values are accessible with much less effort, as the sum over the vibrational states of the excited electronic states can be performed analytically in the harmonic approximation. This also circumvents small errors due to truncation of the vibrational manifold. On the other hand, the full spectra reveal much more detailed information on the role of the vibrational degree of freedom and serve as a more sensitive benchmark when comparing with experiments.

The CD spectrum is defined as the difference in absorption of left and right circularly polarized light \cite{CDnakanishi,schellman_circular_1975}. It is expressed via the respective molar absorption coefficients, $\Delta\epsilon\qty(\omega)=\epsilon_L\qty(\omega) -\epsilon_R\qty(\omega)$ as a function of the frequency of incident electromagnetic radiation.
For a randomly oriented ensemble, considering only electric and magnetic dipole interactions between the molecules and the radiation, resonant excitation with left-circularly and right-circularly polarized light leads in first order perturbation theory to an absorption difference proportional to the imaginary part of the scalar product of the corresponding electric and magnetic transition dipole moments \cite{schellman_circular_1975,meath_differential_1987}. This scalar product constitutes the so-called rotatory strength of the transition \cite{Condon1937}. Therefore, there is a rotatory strength connected to each vibronic transition of the CD spectrum,
\begin{align}
R_{f\vb{v}_f i\vb{0}_i}&=\Im{\vb*{\mu}_{i\vb{v}_i f\vb{0}_f}\cdot\vb{m}_{f\vb{v}_f i\vb{0}_i}}.
\label{eq:R_vibronic}
\end{align}
The connection between the CD spectrum and these quantities is given by \cite{PECUL2005185,schellman_circular_1975}
\begin{widetext}
\begin{align}
  \Delta\epsilon\qty(\omega)&=
  \frac{16\pi^2 N_A \omega}{3\qty(4\pi\epsilon_0)\hbar c^2 \ln{10}}\sum_{\vb{v}_f}
  \rho\qty(\omega,\omega_{f,\vb{v}_f i,\vb{0}_i})R_{f\vb{v}_f i\vb{0}_i} \notag \\
  \flatfrac{\Delta\epsilon\qty(\omega)}{\qty(\mathrm{cm}^2\mathrm{mmol}^{-1})}&=
20.529\times\qty(\flatfrac{\omega}{\qty(E_h\hbar^{-1})})\sum_{\vb{v}_f}
\qty(\flatfrac{\rho\qty(\omega,\omega_{f,\vb{v}_f i,\vb{0}_i})}{\qty(\hbar E_h^{-1})})
\qty(\flatfrac{R_{f\vb{v}_f i\vb{0}_i}}{\qty(e^2a_0\hbar^2 m_e^{-1})}),
\label{eq:deltaeps_vibronic}
\end{align}
\end{widetext}
where $N_A$ is the Avogadro number and $\rho\qty(\omega,\omega_{f,\vb{v}_f i,\vb{0}_i})$ is a normalized line shape function, introducing a line width to each transition $\ket{f \vb{v}_f}\leftarrow\ket{i \vb{0}_i}$ with transition frequency
$\omega_{f,\vb{v}_f i,\vb{0}_i}$.
Since the A-band of the three ketones in our work does not overlap with any other absorption bands it is not necessary to sum over other electronic states in \cref{eq:deltaeps_vibronic}. Note that we adjusted the prefactor in \cref{eq:deltaeps_vibronic} compared to Ref.~\cite{lin_vibronically_2008} to match our definition of the rotatory strength.

The ABS spectrum is defined as the mean absorption of left and right circularly polarized light, $\epsilon\qty(\omega)=\frac{1}{2}\qty(\epsilon_L\qty(\omega)+\epsilon_R\qty(\omega))$ or, equivalently, of unpolarized light. Similar to how the CD spectrum is connected to the rotatory strength, the ABS spectrum is connected to the so-called dipole strength of the involved vibronic transitions, given by the sum of the absolute value squares of electric and magnetic transition dipole moments,
\begin{equation}
D_{f\vb{v}_f i\vb{0}_i}=\abs{\vb*{\mu}_{f\vb{v}_f i\vb{0}_i}}^2+\abs{\vb{m}_{f\vb{v}_f i\vb{0}_i}}^2.
\label{eq:D_vibronic}
\end{equation}
The connection between these quantities is given by \cite{schellman_circular_1975}
\begin{widetext}
\begin{align}
  \epsilon\qty(\omega)&=
  \frac{4\pi^2 N_A \omega}{3\qty(4\pi\epsilon_0)\hbar c \ln{10}}\sum_{\vb{v}_f}
  \rho\qty(\omega,\omega_{f,\vb{v}_f i,\vb{0}_i})D_{f\vb{v}_f i\vb{0}_i} \notag \\
  \flatfrac{\epsilon\qty(\omega)}{\qty(\mathrm{cm}^2\mathrm{mmol}^{-1})}&=
703.301\times\qty(\flatfrac{\omega}{\qty(E_h\hbar^{-1})})\sum_{\vb{v}_f}
\qty(\flatfrac{\rho\qty(\omega,\omega_{f,\vb{v}_f i,\vb{0}_i})}{\qty(\hbar E_h^{-1})})
\qty(\flatfrac{D_{f\vb{v}_f i\vb{0}_i}}{\qty(e^2a_0^2)}).
\label{eq:epsilon_spec}
\end{align}
\end{widetext}
The anisotropy factor spectrum is given as the ratio of CD and ABS, i.e.,
\begin{equation}
g\qty(\omega)=\frac{\Delta \epsilon\qty(\omega)}{\epsilon\qty(\omega)}.
\label{eq:dissymetry_spec}
\end{equation}
Note that if only the Franck--Condon terms are considered and the
line-shape functions for CD and ABS are identical, the anisotropy factor
becomes frequency-independent for a given absorption band, since the
Franck--Condon factors appear as prefactors both in the numerator and denominator of \cref{eq:dissymetry_spec} and thus cancel. 
However, the more complicated structure of the anisotropy seen in experiments suggests the necessity to include Herzberg--Teller effects. As such, to perform adequate calculations of the CD and ABS spectra, the vibrational fine structure of the excited state needs to be taken into account. 

For determining the rotatory strength of the \emph{entire absorption band},
in contrast, the Herzberg--Teller terms do not play a role within our
approximations, which can be seen from the following expression,
\begin{align}
R_{f i}&=\frac{3\qty(4\pi\epsilon_0)\hbar c^2 \ln{10}}{16\pi^2 N_A}\int_{\mathrm{A-band}}\frac{\Delta \epsilon}{\omega}\,\mathrm{d}\omega \notag \\
&=\sum_{\vb{v}_f}
\Im{\vb*{\mu}_{i\vb{0}_i f\vb{v}_f}\cdot\vb{m}_{f\vb{v}_f i\vb{0}_i}}
\approx\Im{\vb*{\mu}_{if}\qty(\vb{0})\cdot\vb{m}_{fi}\qty(\vb{0})}. \label{eq:R_electronic}
\end{align}
Here and in the following, we include FC and linear HT terms for the electric transition dipole moment, whereas we limit ourselves to FC terms in the magnetic transition dipole moment. We used the resolution of the identity via the final state vibrational manifold, $\sum_{\vb{v}_f}\dyad{\vb{v}_f}{\vb{v}_f} = \dblone$, and considered the harmonic approximation in the initial state, which implies that $\mel{\vb{0}_i}{\hat{Q}_r}{\vb{0}_{i}} = 0$ for all $r$.
\Cref{eq:R_electronic} shows that the integrated rotatory strength only depends on the equilibrium structure and is therefore unaffected by Herzberg--Teller contributions. 
Conversely, while the band-integrated dipole strength is unaffected by the final-state vibrational structure, it does depend on the variance of the normal coordinates in the ground state,
\begin{align}
D_{f i}&=\frac{3\qty(4\pi\epsilon_0)\hbar c \ln{10}}{4\pi^2 N_A}\int_{\mathrm{A-band}}\frac{\epsilon}{\omega}\,\mathrm{d}\omega\approx\abs{\vb*{\mu}_{fi}\qty(\vb{0})}^2 \notag \\
&+\sum_r \qty(\abs{\frac{\partial\vb*{\mu}_{fi}}{\partial Q_r}\eval_{\vb{Q}=\vb{0}}}^2\mel{\vb{0}_i}{\hat{Q}^2_r}{\vb{0}_i})+\abs{\vb{m}_{fi}\qty(\vb{0})}^2. \label{eq:D_electronic}
\end{align}
In the harmonic approximation one can directly evaluate $\mel{\vb{0}_i}{\hat{Q}_r^2}{\vb{0}_{i}}=\frac{\hbar}{2\omega_{r,i}}$. 

Using \cref{eq:R_electronic,eq:D_electronic}
we can assign an anisotropy factor to the entire electronic band via \cite{CDnakanishi},
\begin{equation}
g=\frac{4R_{fi}/c}{D_{fi}}.
\label{eq:dissymetry_band}
\end{equation}
As mentioned before, in the FC case, the anisotropy factor spectrum  of \cref{eq:dissymetry_spec} takes on a constant value over the band.
This value then coincides with the band value of \cref{eq:dissymetry_band}, that is $g\qty(\omega)=g$, in this case.
However, in general the anisotropy factor spectrum is not represented by a constant and no such connection can be made \cite{CDnakanishi,schellman_circular_1975}.
Some chiral molecules exist in different conformers at room temperature. For the molecules examined in this work, this is the case for 3MCP. For a mixture of conformers $c$ contained in proportions of $p_{c}$, spectra and integrated band values will be superimposed accordingly, e.g. \cref{eq:dissymetry_spec} becomes in that case,
\begin{equation}
g\qty(\omega)=\frac{\sum_{c}p_{c}\Delta \epsilon_{c}\qty(\omega)}{\sum_{c}p_{c}\epsilon_{c}\qty(\omega)}.
\label{eq:g_electronic_conformers}
\end{equation}
Note that our calculations of the spectral profiles for each individual conformer are still performed for zero temperature, however.

To predict theoretically the integrated quantities of \cref{eq:R_electronic,eq:D_electronic,eq:dissymetry_band}, we first determine the equilibrium
structure of the electronic ground state $\text{S}_0$ and its corresponding vibrational structure within the harmonic approximation.
Then, the transition dipole moments and derivatives of the electric transition dipole moment
are determined.
Determination of the vibrationally resolved spectra represented by \cref{eq:deltaeps_vibronic,eq:epsilon_spec,eq:dissymetry_spec} requires to compute
the Franck--Condon integrals and Herzberg--Teller terms via the equilibrium structure and vibrational structure of the energetically lowest electronically excited singlet state $\text{S}_1$.
For ketones this is complicated by the fact that the
$\text{S}_1$ state of carbonyl compounds typically features two viable minima on its potential energy surface \cite{lin_vibronically_2008}.
This is due to pyramidalization of the carbonylic C atom, which can take place
in two different directions. To account for this, we determined the 
structure of the transition state between both minima,
which corresponds to a nearly planar arrangement around the carbonylic C atom. This stationary point on the $\text{S}_1$ potential energy surface results
in an imaginary vibrational frequency for the mode responsible for pyramidalization.
One could attempt to describe the resulting pyramidalization motion within an effective one-dimensional anharmonic model to capture the corresponding anharmonic progressions in this mode. Another possibility is to choose instead a rather crude and pragmatic model, in which the imaginary vibrational frequency is either replaced by the real frequency of the mode of
the $\text{S}_0$ state with the highest overlap or by its absolute value. 
In this work we chose the latter approach. Moreover, to
determine the 0--0 transition energy we used the adiabatic excitation energy to
the transition structure in S$_1$ and corrected for the difference in the
zero-point vibrational energies with neglection of the imaginary harmonic frequency.
The equilibrium structures and harmonic vibrational force fields of the compounds were determined using
the Turbomole software package \cite{AHLRICHS} (Version 7.6).
Molecular structures were energy-minimized using density functional theory (DFT) at the B3-LYP/def2-TZVP level of theory \cite{B515623H,B508541A,becke,lyp}
with the resolution of the identity method for the Coulomb part (RI-J) \cite{EICHKORN1995} enabled on 
the m4 numerical integration grid with D3 dispersion correction \cite{GrimmeStefanetAl} with damping (BJ) \cite{Grimme2011EffectOT}. More details and results of the quantum chemical calculations are given in the Supplemental Material \cite{supp}.
The vibrational frequencies and normal modes were obtained in the harmonic approximation of the respective potential energy surfaces at their respective structures. Calculation of Franck--Condon and Herzberg--Teller overlaps were performed with a modified version of the hotFCHT code \cite{hotfcht,jankowiak:2007,coriani:2010,huh:2012proc}.

\section{Results}

\begin{figure*}[tb]
\includegraphics[width=1.5\columnwidth]{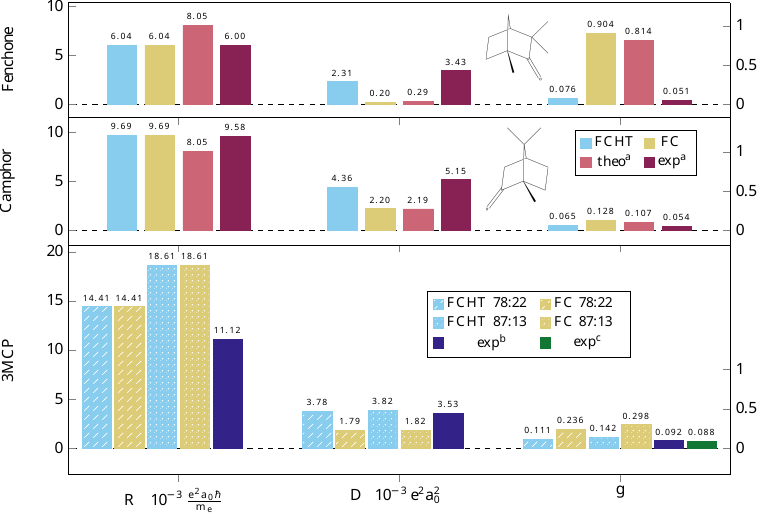}
  \caption{Numerical results for absolute value of rotatory strength, dipole strength (left axis) and absolute value of anisotropy factor (right axis) for fenchone, camphor and 3MCP, including the Franck--Condon and Herzberg--Teller terms for the electric transition dipole moment (FCHT), only including the Franck--Condon term (FC), and comparison with various theoretical \cite{pulm_theoretical_1997}$^a$ and experimental \cite{pulm_theoretical_1997}$^a$,\cite{feinleib_vapour-phase_1968}$^b$,\cite{dekkers_optical_1976}$^c$ values from the literature. For 3MCP, our results are shown for two different mixtures of the equatorial and axial conformers, 87:13 \cite{he_determining_2004} and 78:22 \cite{he_determining_2004,al-basheer_spectroscopic_2007}.}
  \label{fig:integrated_results}
\end{figure*}

\Cref{fig:integrated_results} shows the results of the integrated band values according to \cref{eq:R_electronic,eq:D_electronic,eq:dissymetry_band} for the three ketones (1R,4S)-(-)-fenchone, (1S,4S)-(-)-camphor and (R)-(+)-3-methylcyclopentanone.

To compare our calculations with theoretical and experimental data from the literature we first computed all quantities by only considering 
the Franck--Condon terms (abbreviated as "FC" in the figure). Then, we recalculated all quantities including both the Franck--Condon and Herzberg--Teller terms 
(abbreviated as "FCHT" in the figure). For fenchone and camphor we compare to the experimental and theoretical literature values reported in Ref.~\cite{pulm_theoretical_1997}. 
Note that the theoretical values therein do not take into account HT terms such that we obtain good agreement with our calculations only for the FC case. 
The experimental values differ significantly from the FC results. Specifically, neglecting the HT contribution overestimates the absolute value of the 
anisotropy factor by an order of magnitude for fenchone and a factor of two for camphor. By including these terms the agreement with the experimental values is vastly improved. We attribute this improvement to the large effect of the inclusion of the HT terms on the dipole strengths. The discrepancy between our calculations and the theoretical results from Ref.~\cite{pulm_theoretical_1997} for the rotatory strength is due to differences in the numerical approach.

The ketone 3MCP exists in five conformational forms at room temperature with the equatorial and axial conformers constituting the main components. 
The ratio between these two conformers has been determined using three different approaches in Ref.~\cite{he_determining_2004} yielding proportions of 87:13, 78:22 and 70:30 equatorial:axial 3MCP. 
However, we do not consider the 70:30 ratio in the following, due to the large uncertainty associated with this particular result.
Instead we show results both for a ratio of 87:13, which was obtained via temperature-dependent vibrational absorption intensities (dotted pattern in \cref{fig:integrated_results}),
and for a ratio of 78:22, which was obtained from comparison of theoretically and experimentally determined specific rotations (striped pattern in \cref{fig:integrated_results}). The first ratio coincides with REMPI investigations in the gas phase \cite{kim2000} and the latter ratio 
coincides with a study of conformational proportions in a variety of solvents \cite{al-basheer_spectroscopic_2007}. 
We performed our calculations by first determining the relevant quantities of the equatorial and axial conformers, respectively, and then calculating, e.g., the anisotropy according to \cref{eq:g_electronic_conformers} with the corresponding proportions of the conformers.

\begin{figure*}[tb]
  \includegraphics[width=1.5\columnwidth]{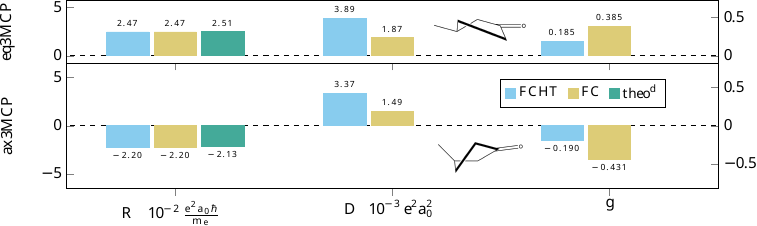}
\caption{Results of rotatory strength, dipole strength (left axis) and anisotropy factor (right axis) for equatorial and axial 3MCP conformers, comparing the values obtained when including both the Franck--Condon and Herzberg--Teller terms for the electric transition dipole moment (FCHT), only the Franck--Condon term (FC), and the rotatory strength obtained in another theoretical study \cite{lin_vibronically_2008}$^d$.}
\label{fig:3mcp_conformers_res}
\end{figure*}

For 3MCP we extracted experimental band-averaged values from Refs.~\cite{feinleib_vapour-phase_1968,dekkers_optical_1976} by digitizing the spectra and calculating the corresponding integrals for the band values.
These experiments encompass both gas phase measurements \cite{feinleib_vapour-phase_1968} and measurements in solution with n-heptane \cite{dekkers_optical_1976}. The values for the band anisotropy factor agree between both experiments.
Comparison with the spectra was complicated by the low resolution in Ref.~\cite{feinleib_vapour-phase_1968}. Additionally, in Ref.~\cite{dekkers_optical_1976} the spectra were only recorded to about $270\,\mathrm{nm}$. 
Nevertheless, these spectra are suitable as a qualitative benchmark for our calculations. \Cref{fig:3mcp_conformers_res} shows the results for the individual conformers of 3MCP.

Since both conformers have similar dipole strengths, but opposite sign in their rotatory strength, the ratio of the mixture between the conformers mainly affects the rotatory strength and the anisotropy.
The inclusion of the HT terms doubles the dipole strengths for both individual conformers and halves the magnitude of the anisotropy 
factors accordingly. Our results for the two conformers agree with a similar theoretical estimate of the rotatory strength from the literature \cite{lin_vibronically_2008}. In that work the authors also included HT contributions 
in the magnetic dipole transition moment, although these turned out to be negligible for the A-band.

\begin{figure}[tb]
\centerline{\includegraphics[width=\columnwidth]{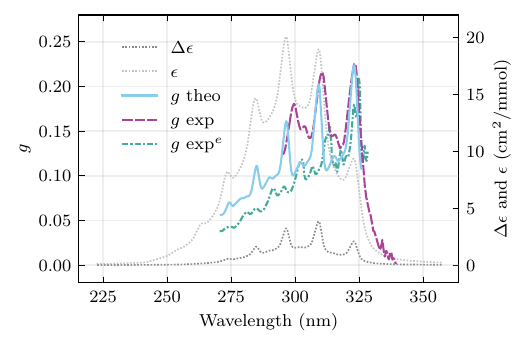}}
\caption{Theoretical anisotropy factor spectrum ($g$, left axis) of 3MCP, assuming a conformational 
    mixture of 78:22 equatorial to axial 3MCP in comparison with our experimental liquid-phase spectrum and an experimental spectrum from the literature \cite{boesl_resonance-enhanced_2013}$^e$. The anisotropy factor spectrum has been generated from the shown CD ($\Delta\epsilon$) and ABS ($\epsilon$) 
	spectra (right axis). A Lorentzian line shape with a FWHM of $0.05\,\mathrm{eV}$ has been applied for CD and ABS spectra. Without HT corrections, the g factor spectrum would correspond to a flat line at $g = 0.236$, see \cref{fig:integrated_results}.}
\label{fig:aniso_spec}
\end{figure}

Our numerically calculated anisotropy factor spectrum for 3MCP is shown as a solid blue line in \cref{fig:aniso_spec}.
Here and in the following we only show our theoretical results inside the A band, omitting the data points where the A band absorption approaches zero as this regime is not well-captured by our model.
The spectrum was generated assuming a conformational mixture of 78:22 equatorial to axial 3MCP \cite{he_determining_2004,al-basheer_spectroscopic_2007}. For the line shape function, we employed a Lorentzian profile of $0.05\,\mathrm{eV}$ full width half maximum, similarly to Ref.~\cite{lin_vibronically_2008}. 
We neglect the magnetic contributions in \cref{eq:D_vibronic} since they turned out to be vanishingly small in our calculations.

We furthermore compare our calculated anisotropy factor spectrum with our experimentally obtained liquid-phase spectrum of 3MCP.
The respective absorption studies were solely performed on the
(R)-enantiomer (Sigma Aldrich, constitutional purity 99.8\%). The
substance was used without further purification and dissolved in the
non-polar solvent n-hexane. We employed a circular dichroism
spectropolarimeter (J-815, Jasco) to measure the relative absorbance of
a solution of $65\,\mathrm{mL}$ 3-methylcyclopentanone in $935\,\mathrm{mL}$ n-hexane. An
average of six consecutive measurements with step size and bandwidth of
$0.5\,\mathrm{nm}$ is considered. The background by the solvent was determined
separately and subtracted. A UV/Visible spectrophotometer (Cary 100,
Varian) with simultaneous acquisition of background and sample
absorbance was used to evaluate the absolute absorption. The step size
was $0.5\,\mathrm{nm}$, and the bandwidth was increased to $1\,\mathrm{nm}$. Here, $20\,\mathrm{mL}$ 3MCP
were dissolved in $10\,\mathrm{mL}$ n-hexane for good signal-to-noise-ratio. In the
region close to $200\,\mathrm{nm}$, the cutoffs of the quartz cuvettes containing
the samples and n-hexane are reached, i.e., their absorption increases already
for larger wavelengths. In addition, the CD data from the first
measurement becomes noisy in this wavelength region. The next strong
absorption band with pronounced CD is located in the wavelength region
from $200\,\mathrm{nm}$ to $185\,\mathrm{nm}$, according to Ref.~\cite{feinleib_vapour-phase_1968}, but cannot
be observed under the present measurement conditions.
Compared to the
gas phase \citep{boesl_resonance-enhanced_2013}, the liquid-phase measurement using the non-polar
solvent n-hexane only shows a soft hypsochromic shift ($\sim 0.5\,\mathrm{nm}$).

Our calculations reproduce the most important features of the experimental spectra very well. Specifically, the sign and overall magnitude of the anisotropy factor is well-reproduced and the position of the main peaks at 
around $300\,\mathrm{nm}$, $310\,\mathrm{nm}$ and $320\,\mathrm{nm}$ matches. Moreover, the peak heights increase from smaller to larger wavelengths with several smaller
peaks interspersed. We thus expect our calculations to capture the main influence of the vibronic structure of 3MCP on the anisotropy factor. 

\begin{figure}[tb]
\centerline{\includegraphics[width=\columnwidth]{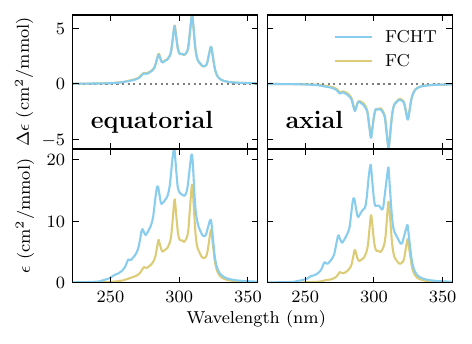}}
\caption{CD (upper plots) and ABS (lower plots) spectra of equatorial and axial 3MCP. The plots compare the results for FC and FCHT calculations. A Lorentzian line shape with a FWHM of $0.05\,\mathrm{eV}$ has been employed.}
\label{fig:conformers_cd_abs_specs}
\end{figure}

\Cref{fig:conformers_cd_abs_specs} unravels the source of the strong deviation from a constant anisotropy factor spectrum for 3MCP which would be expected if the absorption were purely Franck--Condon dominated.
CD and ABS spectra are shown for both the axial and equatorial conformers. The spectral position of the A-band for the axial conformer is found to be at $0.017\,\mathrm{eV}$ below the spectral position for the equatorial conformer. This coincides with experimental findings for the energy shift of the B-band \cite{kim2000}.
Similarly to the band-integrated data in \cref{fig:integrated_results}, we show our results once obtained with and without the inclusion of HT terms.

For the spectra of equatorial 3MCP as well as the conformational mixture, the main peaks are approximately located at $323\,\mathrm{nm}$, $309\,\mathrm{nm}$, $296\,\mathrm{nm}$, $284\,\mathrm{nm}$, and  $273\,\mathrm{nm}$.
The peak separation is about $0.17\,\mathrm{eV}$, corresponding to the energy of the C-O stretching mode.
A detailed discussion of the vibrational structure has been given elsewhere \cite{lin_vibronically_2008,boesl_resonance-enhanced_2013}. In agreement with other theoretical studies \cite{lin_vibronically_2008}, our results show that Herzberg--Teller 
contributions do not affect the CD spectrum in the case of 3MCP.
A possible explanation was given in the literature \cite{lin_vibronically_2008}.
Therein, a lack of energetically close large intensity transitions
was considered to prevent a large intensity borrowing effect.
However, this reasoning has been questioned before \cite{orlandi_theory_1973}
and, in fact, does not appear to provide an adequate explanation to our results.
Specifically, as evidenced by the bottom panels of \cref{fig:conformers_cd_abs_specs} the Herzberg--Teller effect
significantly alters the ABS spectrum and therefore explains the strong dependence of anisotropy factors on the vibrational structure found in experiments.

\begin{figure}[tb]
\centerline{\includegraphics[width=\columnwidth]{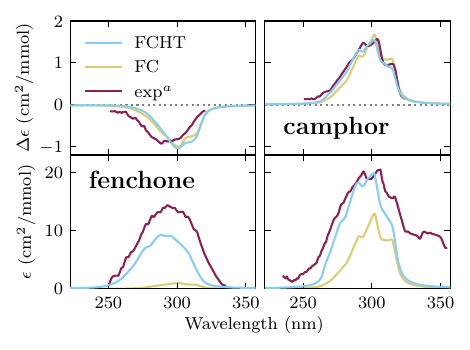}}
\caption{CD (upper plots) and ABS (lower plots) spectra of fenchone and camphor. The plots compare the results for FC and FCHT calculations. A Lorentzian line shape with a FWHM of $0.08\,\mathrm{eV}$ has been employed.}
\label{fig:cd_abs_specs_fenchone_camphor}
\end{figure}

\begin{figure}[tb]
\centerline{\includegraphics[width=\columnwidth]{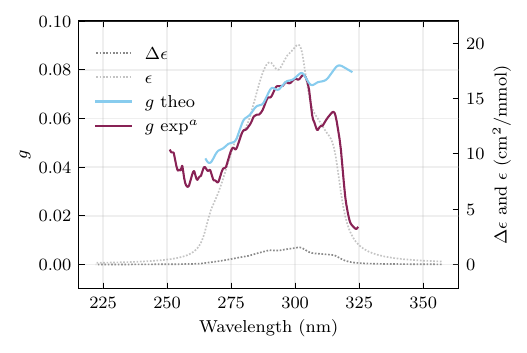}}
    \caption{Theoretical and experimental \cite{pulm_theoretical_1997}$^a$ anisotropy factor spectrum ($g$, left axis) of camphor. The anisotropy factor spectrum has been generated from the shown CD ($\Delta\epsilon$) and ABS ($\epsilon$) spectra (right axis). A Lorentzian line shape with a FWHM of $0.08\,\mathrm{eV}$ has been employed for CD and ABS spectra. Note that we flipped the sign for CD and anisotropy factor to match the enantiomer examined in the experiment. Without HT corrections, the g factor spectrum would correspond to a flat line at $g = 0.128$, see \cref{fig:integrated_results}.}
\label{fig:camphor_aniso}
\end{figure}

For fenchone and camphor, we chose a Lorentzian profile of $0.08\,\mathrm{eV}$ full width half maximum for the line shape function.
This choice was made in view of obtaining good agreement with the experimental spectra of Ref.~\cite{pulm_theoretical_1997}.
The inclusion of the Herzberg--Teller terms leads to some deformation of the CD spectrum when compared to the Franck--Condon results shown in \cref{fig:cd_abs_specs_fenchone_camphor}. However, on a qualitative level the curves do not differ appreciably with a significant impact of the Herzberg--Teller terms only visible for the ABS spectra.
Our theoretical results for camphor show remarkable agreement with the experimental data \cite{pulm_theoretical_1997}, which can also be seen in the corresponding anisotropy factor spectrum in \cref{fig:camphor_aniso}.

\begin{figure}[tb]
\centerline{\includegraphics[width=\columnwidth]{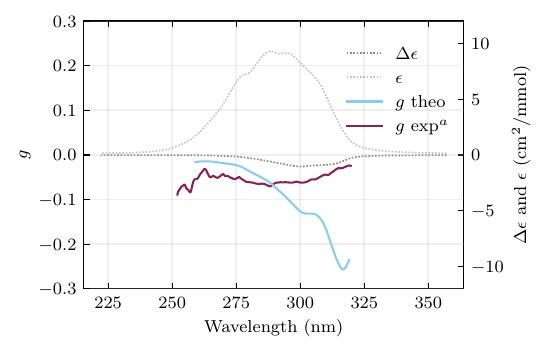}}
	\caption{Theoretical and experimental \cite{pulm_theoretical_1997}$^a$ anisotropy factor spectrum ($g$, left axis) of fenchone. The anisotropy factor spectrum has been generated from the shown CD ($\Delta\epsilon$) and ABS ($\epsilon$) spectra (right axis). A Lorentzian line shape with a FWHM of $0.08\,\mathrm{eV}$ has been applied for CD and ABS spectra. Without HT corrections, the g factor spectrum would correspond to a flat line at $g = 0.904$, see \cref{fig:integrated_results}.}
\label{fig:fenchone_aniso}
\end{figure}

Conversely, for fenchone our results in \cref{fig:cd_abs_specs_fenchone_camphor} show that the theoretical spectra are shifted slightly in energy compared to the experimental data. Moreover, the shapes of the spectral curves do only agree well for the CD spectrum. Nonetheless, we can see a strong improvement towards agreement with the experimental results for the ABS spectrum through inclusion of HT terms. Accordingly, the anisotropy factor spectrum shown in \cref{fig:fenchone_aniso} does not quantitatively agree with the experiment.

We attribute the remaining discrepancies between our calculations and the experimental data to several potential causes.
First, there is an inherent limited accuracy in our quantum chemical calculations. This concerns especially equilibrium structures, transition dipole moments and their derivatives with respect to displacements along the normal modes, as well as transition energies. 
This is particularly relevant for 3MCP as the spectra are constituted from two conformers. 
Second, involving the conformers of 3MCP, also their ratio, which is not exactly known, is particularly relevant to the CD spectra and the rotatory strength.
Third, temperature effects, most notably leading to a Boltzmann distribution of the initial population in excited vibrational levels in the electronic ground state could affect the spectra and extend them to lower energies.
Fourth, the quantum chemical calculations performed in this work do not consider any interactions between the molecules or between the molecules and a solvent, which limits our ability to reproduce experimental results recorded in the liquid phase. This is is particularly relevant for 3MCP, as both conformers can be affected differently by the solvent.
Lastly, the approximation of the potential energy surfaces as harmonic potentials is a strong simplification for the double well potential of the excited electronic states in the ketones we studied.

\section{Conclusion}

We have calculated the absorption and anisotropy factor spectra for the A-band transition in the three ketones (1R,4S)-(-)-fenchone, (1S,4S)-(-)-camphor and (R)-(+)-3-methylcyclopentanone. To this end we employed quantum chemistry calculations at the DFT level in the harmonic approximation to determine vibrational frequencies and normal modes in the corresponding electronic states. By including the Herzberg--Teller contribution in the vibronic matrix elements for the electric transition dipole moments we obtained both frequency-resolved absorption and anisotropy factor spectra as well as the corresponding band-integrated results.
Even though their impact on the difference in absorption between the enantiomers is small, we found that the Herzberg--Teller contributions are crucial to understand the structure of the absorption spectra and therefore also the anisotropy.
Furthermore, when only considering the Franck--Condon contributions the band-integrated value of the anisotropy can differ by up to an order of magnitude as evidenced by fenchone. This highlights the critical importance of the Herzberg--Teller effect when investigating circular dichroism in the approximately electric dipole-forbidden transitions for which particular large dichroic signatures can be observed in absorption.

Our calculations show good agreement with experimental data for all three ketones, particularly for the band-integrated values.
For 3MCP we are able to reproduce many central features of the experimentally reported anistropy spectra. Although our agreement for fenchone is weaker we are still able to reproduce the general magntiude and most important qualitative features.

Our results pave the way for an improved modeling of circular dichroism experiments in ketones. This is of particular importance for understanding the role of the vibrational degree of freedom on the anisotropy when using shaped, e.g. chirped, laser pulses.
The interplay between the chirp and the vibrational structure can be an important resource for coherent control as has been evidenced for example in the photoassociation of Mg$_2$ dimers \cite{LevinPRL15,LevinJPhysB2021}. 
Moreover, the improved modeling of the vibronic transitions will be an important asset to lift studies on optimal control of circular dichroism from few-level systems closer to physical reality \cite{mondelo-martell_increasing_2022}.
Finally, to understand the difference in anisotropy between absorption and ion yield we expect that further theoretical investigations in the role of higher-lying electronic excited states are required which will be subject to future work.

\begin{acknowledgments}
We thank the AG Biophysik by Prof. Kleinschmidt at Universit\"{a}t Kassel for their support in recording the anisotropy factor spectrum for 3MCP. Financial support by the
Deutsche Forschungsgemeinschaft (DFG, German Research Foundation)—Projektnummer 328961117—SFB ELCH 1319 is gratefully acknowledged.
\end{acknowledgments}

\bibliography{./anisotropy.bib}

\end{document}